%% file: main.tex
\pdfoutput=1
\documentclass[conference,a4paper,10pt]{IEEEtran}
\usepackage[latin1]{inputenc}
\usepackage{times,amsmath}
\usepackage{amsfonts}
\usepackage{pstool}
\usepackage{subfigure}
\usepackage{multirow}
\usepackage{enumerate}
\usepackage{graphicx}
\usepackage{MnSymbol}
\usepackage{stfloats} 
\usepackage[table]{xcolor}
\usepackage[square, comma, sort&compress, numbers]{natbib}
\usepackage{nohyperref}
\usepackage{algorithm,algorithmic}
\usepackage{bigints}
\usepackage{epstopdf}

\newcounter{tempEquationCounter} 
\newcounter{thisEquationNumber}

\makeatletter
\newcommand{\vast}{\bBigg@{4}}
\newcommand{\Vast}{\bBigg@{5}}
\newcommand{\norm}[1]{\left\lVert#1\right\rVert}
\makeatother

\graphicspath{ {Figures/} }

\begin{document}

\title{Exploiting Lack of Hardware Reciprocity for Sender-Node Authentication at the PHY Layer}

\author{
\IEEEauthorblockN{Muhammad Mahboob Ur Rahman\IEEEauthorrefmark{1}, Aneela Yasmeen\IEEEauthorrefmark{2}}
\IEEEauthorblockA{\IEEEauthorrefmark{1}Electrical engineering department, Information Technology University (ITU), Lahore, Pakistan \\ mahboob.rahman@itu.edu.pk }
\IEEEauthorblockA{\IEEEauthorrefmark{2}Department of Computer and Systems Sciences, Stockholm university, Stockholm, Sweden \\ u04181@yahoo.com }
}

\maketitle

\input{abstract_reci}

\input{sec1_reci_intro}

\input{sec2_reci_SM}

\input{sec3_reci}
\input{sec4_reci}

\input{sec5_reci}

\input{sec6_reci_sim}
\input{conclusion_reci}

\appendices

\footnotesize{
\bibliographystyle{IEEEtran}
\bibliography{references}
}

\vfill\break

\end{document}

%% file: abstract_reci.tex
\begin{abstract} 

This paper proposes to exploit the so-called {\it  reciprocity parameters} (modelling non-reciprocal communication hardware) to use them as decision metric for binary hypothesis testing based authentication framework at a receiver node Bob. Specifically, Bob first learns the reciprocity parameters of the legitimate sender Alice via initial training. Then, during the test phase, Bob first obtains a measurement of reciprocity parameters of channel occupier (Alice, or, the intruder Eve). Then, with ground truth and current measurement both in hand,  Bob carries out the hypothesis testing to automatically accept (reject) the packets sent by Alice (Eve). For the proposed scheme, we provide its success rate (the detection probability of Eve), and its performance comparison with other schemes.

\end{abstract}

%% file: sec1_reci_intro.tex
\section{Introduction}
\label{sec:intro}

Physical-layer authentication is a well-studied problem within the domain of Physical-layer security. There, the task of a receiver node {\it Bob} is to exploit some attribute of the physical layer (wireless medium or communication hardware) to use it as sender's fingerprint in order to accept (reject) the packets coming from the legitimate sender {\it Alice} (the intruder {\it Eve}) systematically. So far, researchers have considered channel frequency response (CFR) \cite{Xiao:TWC:2008}, channel impulse response (CIR) \cite{Jitendra:COMSNETS:2010},\cite{Wang:ICC:2013}, carrier frequency offset (CFO) \cite{Mahboob:ICUWB:2015},\cite{Mahboob:Globecom:2014}, angle-of-arrival (AoA) \cite{Xiong:2010:SIGCOMM}, IQ-imbalance \cite{Wang:ICC:2014}, received signal strength (RSS) \cite{Trappe:TPDS:2013} etc. to use them as sender's fingerprints for authentication. 

The schemes proposed in \cite{Xiao:TWC:2008}-\cite{Trappe:TPDS:2013} all share a common framework for authentication. That is, Bob first acquires the ground truth via training with Alice on a secure channel; then later during the test phase, Bob authenticates every packet received from the shared channel by doing hypothesis testing on current measurement of sender's fingerprint against the ground truth. Table 1, on last page, provides a qualitative comparison of schemes in \cite{Xiao:TWC:2008}-\cite{Trappe:TPDS:2013} as well as the proposed scheme.

At this point, it is worth mentioning that the proposed reciprocity-based authentication method conceptually differs from the RF fingerprinting based authentication methods (e.g., \cite{Goeckel:JSAC:2011}) as well as channel based authentication methods (e.g., \cite{Xiao:TWC:2008}). Specially, RF fingerprinting based methods (e.g., \cite{Goeckel:JSAC:2011}) exploit non-ideal characteristics of individual components of communication hardware, e.g., ADC, power amplifiers etc. Contrary to such methods, the proposed method neither measures nor exploits the individual values of reciprocity parameters. Rather the proposed method carries out two-way message exchange between a node pair to compute the so-called {\it residual channel} (which is non-zero due to the fact that communication hardware is not reciprocal even when the radio channel is); the residual channel then acts as transmitter fingerprint. Next, even though the proposed method measures the device-to-device channel in both directions (while channel based methods, e.g., \cite{Xiao:TWC:2008}, record the channel frequency/impulse response in one direction only), the RF channels in both directions cancel each other due to reciprocity; therefore, the actual transmitter fingerprint utilized by the proposed method is the residual channel, and not the RF channel itself.

{\bf Contributions.} The main contributions of this paper are the following: i) proposal as well as algorithmic solution to exploit {\it reciprocity parameters} as device fingerprint for sender-node authentication, ii) performance analysis of proposed scheme (i.e., probability of detection of Eve). Furthermore, the proposed {\it two-way} message exchange protocol, the so-called ping-pong iteration (see section III-A), for sender's device fingerprint acquisition lends the proposed method readily available for integration into challenge-response based Authorization systems \cite{Davis:USPatent:2000}. Additionally, the proposed method also finds its application in transmitter identification problem \cite{Goeckel:JSAC:2011}, intrusion detection problem \cite{Onat:WiMob:2005} and two-way authentication problem.  

{\bf Notations.} $(.)^*$ denotes complex-conjugate operation; $(.)^H$ denotes the hermitian-transpose operation; $\norm{.}$ denotes the $2$-norm; $\mathbf{I}_K$ denotes a size $K\times K$ identity matrix; $\mathbb{E}(.)$ denotes the expectation operator.

%% file: sec2_reci_SM.tex
\section{System Model and Background}
\label{sec:sys-model}

\subsection{System Model}
\begin{figure}[ht]
\begin{center}
	\includegraphics[width=2in]{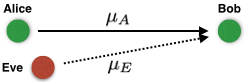} 
\caption{System model}
\label{fig:sys-model}
\end{center}
\end{figure}

We consider a narrow-band, time division duplex (TDD) system where the sender nodes transmit on a shared, time-slotted, block fading channel. Specifically, Alice and Bob make a legitimate transmit-receive pair whereas Eve is an active intruder whose objective is to impersonate Alice (see Fig. \ref{fig:sys-model}). In other words, whenever Alice is absent, Eve sends malicious data to Bob and strives to make Bob believe that she is indeed Alice. Inline with previous literature \cite{Xiao:TWC:2008}-\cite{Trappe:TPDS:2013}, we assume the following: i) Eve is a strong adversary who could easily learn about presence/absence of Alice (e.g., by means of spectrum sensing) in the beginning of every time-slot; ii) Eve, being a clever impersonator (and not a mere jammer), tends to avoid collisions on the shared channel so as to stay undetected.

\subsection{Background: Non-reciprocal Communication Hardware}

{\it Reciprocity parameters} (RP) model the non-reciprocal nature of the factory-manufactured wireless communication hardware; and therefore, are device-dependent and unique. There are two RPs per device, one for each of the RF chains \cite{Guillaud:SPA:2005}. RPs arise due to different amount of magnitude and phase distortion caused by the RF Tx chain and RF Rx chain of a communication device. RPs are relatively time-invariant due to their temperature-dependent nature. To date, there is no known wireless device which comes with reciprocal hardware \cite{Guillaud:SPA:2005}. It is then the non-reciprocal hardware which makes the {\it device-to-device} channel non-reciprocal, even when the radio channel itself is reciprocal.

%% file: sec3_reci.tex
\section{The Proposed Method}
\label{sec:methods}

The proposed method consists of two steps: i) acquisition of sender's device fingerprint, ii) hypothesis testing for authentication. Additionally, whenever the reciprocity parameters change, Bob updates its estimate of ground truth (aka fingerprint of Alice) via training with Alice on a secure channel. 

\subsection{Acquisition of Sender's Device Fingerprint}

This step consists of a ping-pong iteration between Bob and sender node followed by least squares (LS) estimation of sender's device fingerprint.

\subsubsection{Ping-Pong Iteration between Bob and Sender Node}

\begin{figure}[ht]
\begin{center}
	\includegraphics[width=2.8in]{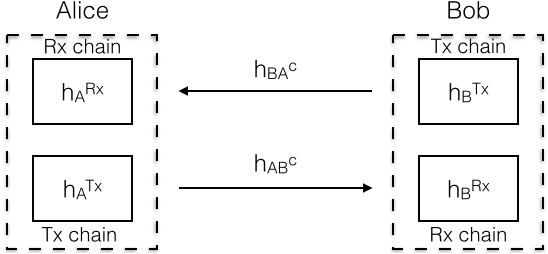} 
\caption{Ping-pong message exchange between Bob and Alice.}
\label{fig:pingpong}
\end{center}
\end{figure}

During every time-slot, Bob broadcasts a {\it ping} preamble $x_B$ with power $P_B$ on the shared channel. This ping message could be the response to an earlier {\it channel access request} by the sender node (Alice or Eve). Then, the signal $y_S$ received by the sender node $S$ ($S \in \{A\equiv Alice,E\equiv Eve\}$) is:
\begin{equation}
	\label{eq:yA}
	 y_S = \sqrt{P_B} h_{BS}. x_B + n_S    
\end{equation}
where $n_S \sim \mathcal{CN}(0,\sigma_{S}^2)$ is the noise at the sender $S$ and
\begin{equation}
	\label{eq:hBA}
	 h_{BS} = h_B^{Tx} . h_{BS}^c . h_S^{Rx} . \exp{\{j(2\pi f_{BS}t+\phi_{BS})\}}    
\end{equation}
is the effective {\it directional} channel from Bob to $S$ \cite{Guillaud:SPA:2005}. $h_{BS}$ includes the radio channel $h_{BS}^c$, reciprocity parameters $h_B^{Tx}$, $h_S^{Rx}$ of Bob and $S$ (see Fig. \ref{fig:pingpong}), and frequency and phase offsets $f_{BS}$, $\phi_{BS}$ (due to oscillators' mismatch). Assuming that $h_{BS}^c \sim \mathcal{CN}(0,1)$, we get $h_{BS} \sim \mathcal{CN}(0,|h_B^{Tx}.h_S^{Rx}|^2)$.

Next, $S$ immediately echoes-back the received signal $y_S$ as a {\it pong} message to Bob using {\it amplify-and-forward} (AF) relaying\footnote{The pong response by Eve (to ping message by Bob) during the test phase should not be considered as an act of cooperation by Eve. Rather, it is part of the proposed protocol; therefore, Eve must abide by the protocol if she wants to intrude into the system while staying undetected. Such two-way message exchange is common in cryptography-based Authorization systems \cite{Davis:USPatent:2000}.}. This ensures that the ping-pong iteration takes place in a time interval much shorter than the channel coherence interval for the given environment. Then, $z_{B|S}$, to be read as "signal received by Bob due to transmission by sender $S$", is given as:
\begin{equation}
	\label{eq:zB}
	z_{B|S} = \beta_{SB}. h_{SB}. y_S + n_B    
\end{equation}
where $n_B \sim \mathcal{CN}(0,\sigma_{B}^2)$ is the noise at Bob and
\begin{equation}
	\label{eq:hAB}
	 h_{SB} = h_S^{Tx} . h_{SB}^c . h_B^{Rx} . \exp{\{j(2\pi f_{SB}t+\phi_{SB})\}}    
\end{equation}
is the effective directional channel from $S$ to Bob (see Fig. \ref{fig:pingpong}). Assuming that $h_{SB}^c \sim \mathcal{CN}(0,1)$, we get $h_{SB} \sim \mathcal{CN}(0,|h_S^{Tx}.h_B^{Rx}|^2)$. Moreover, $\beta_{SB}$ is a scaling factor used by $S$ so as to satisfy the transmit power constraint:
\begin{equation}
	\label{eq:beta}
	 \beta_{SB} = \sqrt{\frac{P_S}{P_B.\ |h_{BS}|^2 +\sigma_{S}^2}}    
\end{equation}

We first note that $h_{SB}^c = (h_{BS}^c)^*$, $f_{SB} = -f_{BS}, \phi_{SB} = -\phi_{BS}$ (assuming negligible oscillator drift during the ping-pong iteration). Then, plugging (\ref{eq:yA}),(\ref{eq:hBA}),(\ref{eq:hAB}) into (\ref{eq:zB}) yields:
\begin{equation}
	\label{eq:zB_simp}
	 z_{B|S} = \sqrt{P_B}. \beta_{SB}. h_S^{Tx} . h_B^{Rx} . h_B^{Tx} . h_S^{Rx} . x_B + {n}_{B|S}    
\end{equation}
where ${n}_{B|S} = \beta_{SB} .h_{SB}.n_S + n_B$ is the net noise at Bob; ${n}_{B|S} \sim \mathcal{CN}(0,{\sigma}_{{B|S}}^2)$ where ${\sigma}_{{B|S}}^2 = \beta_{SB}^2.|h_{SB}|^2.\sigma_{S}^2 + \sigma_{B}^2 $. Let  $\tilde{h}_{SB} = h_S^{Tx} . h_B^{Rx} . h_B^{Tx} . h_S^{Rx} $. Then, $\tilde{h}_{SB}$ is the so-called {\it residual channel}, basically a complex scalar which contains all the four reciprocity parameters (between Bob and $S$). In the proposed method, $\tilde{h}_{SB}$ serves as device fingerprint of $S$; therefore, Bob needs to estimate $\tilde{h}_{SB}$ from $z_{B|S}$. Then:
\begin{equation}
	\label{eq:zB_AF}
	 z_{B|S}^{AF} = \sqrt{P_B}. \beta_{SB}. \tilde{h}_{SB} . x_B + {n}_{B|S}    
\end{equation}

When $S$ employs {\it decode-and-forward} (DF) relaying, it constructs the pong message by pre-multiplying the known $x_B$\footnote{The preamble $x_B$ is not known (known) to sender nodes, i.e., Alice and Eve, when they do AF (DF) relaying to generate the pong message.} with $h_{BS}$ (perfectly) estimated from (\ref{eq:yA}) and applies the gain $\sqrt{P_S}$. Bob then receives the following: 
\begin{equation}
	\label{eq:zB_DF}
	 z_{B|S}^{DF} = \sqrt{P_S}. \tilde{h}_{SB} . x_B + n_B    
\end{equation}

\subsubsection{Least Squares Estimation of Sender's Fingerprint}
In order to estimate the fingerprint $\tilde{h}_{SB}$ of $S$, Bob sends out $K$ training symbols $\mathbf{x_B} = [x_B^1,..., x_B^K]^T$ in the ping preamble; $S$ then echoes-back (via AF/DF relaying) with a pong message. Then, we rewrite (\ref{eq:zB_AF}),(\ref{eq:zB_DF}) in vector form as:
\begin{equation}
	\label{eq:zBvec_AF}
	 \mathbf{z}_{B|S}^{AF} = \sqrt{P_B}. \beta_{SB}. {\tilde{h}}_{SB}. \mathbf{x}_B + \mathbf{{n}}_{B|S}    
\end{equation}
\begin{equation}
	\label{eq:zBvec_DF}
	 \mathbf{z}_{B|S}^{DF} = \sqrt{P_S}. {\tilde{h}}_{SB}. \mathbf{x}_B + \mathbf{n}_B    
\end{equation}
where $\mathbf{{n}}_{B|S} \sim \mathcal{CN}(\mathbf{0},{\sigma}_{n_{B|S}}^2\mathbf{I}_K)$. Let $\mathbf{\tilde{x}}_{B|S}^{AF} = \sqrt{P_B}. \beta_{SB}.\mathbf{x}_B$, $\mathbf{\tilde{x}}_{B|S}^{DF} = \sqrt{P_S}. \mathbf{x}_B$. Then, Bob obtains the two LS estimates: 
\begin{equation}
	\label{eq:Esth}
	 \overline{ \tilde{h}_{SB}^{AF} } = \frac{(\mathbf{\tilde{x}}_{B|S}^{AF})^H \mathbf{z}_{B|S}^{AF}}{\norm{ \mathbf{\tilde{x}}_{B|S}^{AF} }^2};
          \overline{ \tilde{h}_{SB}^{DF} } = \frac{(\mathbf{\tilde{x}}_{B|S}^{DF})^H \mathbf{z}_{B|S}^{DF}}{\norm{ \mathbf{\tilde{x}}_{B|S}^{DF} }^2}  
\end{equation}
where $\overline{ \tilde{h}_{SB}^{AF} } \sim \mathcal{CN}({\tilde{h}}_{SB}, \Sigma_{B|S}^{AF})$, $\overline{ \tilde{h}_{SB}^{DF} } \sim \mathcal{CN}({\tilde{h}}_{SB}, \Sigma_{B|S}^{DF})$ with $\Sigma_{B|S}^{AF} = \frac{{\sigma}_{{B|S}}^2}{K P_B \beta_{SB}^2}$ and $\Sigma_{B|S}^{DF} = \frac{{\sigma}_{{B}}^2}{K P_S}$.

\subsection{Hypothesis Testing for Authentication}

Let $p^R \doteq \overline{ \tilde{h}_{SB}^R }$ ($R \in \{AF,DF\}$), $\mu_A\doteq \tilde{h}_{AB}$ and $\mu_E\doteq \tilde{h}_{EB}$. Bob utilizes the LS estimate of sender's fingerprint from (\ref{eq:Esth}) to cast the sender-node authentication problem as a binary hypothesis testing problem:

\begin{equation}
	\label{eq:H0H1}
	 \begin{cases} H_0: & {p}^R = \tilde{h}_{AB} + {\epsilon_{B|A}^{R}} \\ 
                                 H_1: & {p}^R = \tilde{h}_{EB} + {\epsilon_{B|E}^{R}} \end{cases}
\end{equation}
where ${\epsilon_{B|S}^R}\sim \mathcal{CN} ({0}, \Sigma_{B|S}^R)$ is the estimation error. Then ${p}^R|H_0 \sim \mathcal{CN}(\mu_{A}, \Sigma_{B|A}^R)$ and ${p}^R|H_1 \sim \mathcal{CN}(\mu_{E}, \Sigma_{B|E}^R)$. If $H_0=1$ ($H_1=1$), received data is accepted (rejected).

Next, since $\mu_A$ is available (due to initial training), Bob applies the following test:
\begin{equation}
	\label{eq:H0H1_1_2}
	 T^R = | p^R - \mu_A | \underset{H_0}{\overset{H_1}{\gtrless}} \delta^R   
\end{equation}
where $\delta^R$, the decision threshold, is a design parameter.

%% file: sec4_reci.tex
\section{Implementation of the proposed method}
\label{sec:perf}

Implementation of the hypothesis test in (\ref{eq:H0H1_1_2}) requires a suitably chosen value of the design parameter $\delta^R$. This work computes $\delta^R$ by following the Neyman-Pearson procedure, i.e., the probability of false alarm $P_{fa}$ is set to a desired (error tolerance) value to compute $\delta^R$. More precisely, let $q^R = p^R - \mu_A $. Then, $q^R|H_0 \sim \mathcal{CN} ({0}, \Sigma_{B|A}^R)$. Then, the test statistic $T^R|H_0 \sim Rayleigh(\rho=\sqrt{{\Sigma_{B|A}^R}/{2}})$. Then:
\begin{equation} \label{eq:pfa_1}
	P_{fa} = Pr(|q^R|>\delta^R |H_0) = \exp{\bigg(\frac{-(\delta^R)^2}{\Sigma_{B|A}^R}\bigg)}
\end{equation}

By setting $P_{fa}$ to a pre-specified value, $\delta^R$ is calculated as:
\begin{equation}
	\label{eq:delta}
	\delta^R = \sqrt{-(\ln(P_{fa})) (\Sigma_{B|A}^R)} 
\end{equation}

{\bf Corollary 1.} When sender nodes do AF relaying, $\delta^{AF}\doteq Y$ in (\ref{eq:delta}) is then a random variable with probability density function: $f_{Y}(y,\lambda,c)=2\lambda. y \exp{ \{-\lambda(y^2-c)\} }$; $y\ge \sqrt{c}$ (which is quite similar to Rayleigh distribution); where $\lambda=\frac{K.P_B}{-\ln(P_{fa}).\sigma_A^2.\bar{\gamma}_{AB}}$ where $\bar{\gamma}_{AB}$ is the average SNR of link between Alice and Bob, and $c=\frac{-\ln(P_{fa}).\sigma_B^2}{K.P_B.\beta_{AB}^2}$. Then:
\begin{equation}
	\label{eq:delta_approx}
	\hat{\delta}^{AF} =\mathbb{E}(\delta^{AF})= \frac{ \frac{1}{\bar{\gamma}_{AB}}. (\frac{\sigma_B^2}{\sigma_A^2.\beta_{AB}^2})+1 }{ \frac{1}{\bar{\gamma}_{AB}}. (\frac{K.P_B}{-\ln(P_{fa}).\sigma_A^2}) } 
\end{equation}

{\bf Remark 1.} (\ref{eq:delta}) signifies that the computation of $\delta^R$ requires knowledge of the variance $\Sigma_{B|A}^R$. Then, following two distinct cases are visible: i) when sender nodes employ DF relaying to generate pong message, $\Sigma_{B|A}^{DF}$ is deterministic and known to Bob, then (\ref{eq:delta}) holds; ii) when sender nodes employ AF relaying, either Bob knows $h_{AB}$ (and hence $\Sigma_{B|A}^{AF}$), then (\ref{eq:delta}) holds; or, Bob knows only the distribution of $\Sigma_{B|A}^{AF}$, then, Bob could make the following approximation: $\delta^{AF}\approx \hat{\delta}^{AF}$ (i.e., Bob substitutes $\hat{\delta}^{AF}$ from (\ref{eq:delta_approx}) as ${\delta}^{AF}$ in (\ref{eq:H0H1_1_2})). In short, for the proposed method to work, only knowledge about the channel/pathloss between Alice and Bob is required. 

The proposed method is summarized in Algorithm \ref{alg:1}.

\begin{algorithm}
\caption{The proposed method.}
\label{alg:1}

\begin{algorithmic}
\STATE \textbf {Phase-I: training} \textcolor{blue}{//$H_0=1$ }
\STATE Bob does one or more ping-pong iterations with Alice to estimate the ground truth $\mu_A$ via (\ref{eq:Esth}).  
\STATE \textbf {Phase-II: Authentication} \textcolor{blue}{//Done every time-slot. }
\STATE 1) Bob does one ping-pong iteration with channel occupant to compute the current measurement $p^R$ via (\ref{eq:Esth}).
\STATE 2) Bob computes the threshold $\delta^R$ via (\ref{eq:delta}) or (\ref{eq:delta_approx}).
\STATE 3) Bob implements the test in (\ref{eq:H0H1_1_2}) to accept/reject packets.
\STATE \textcolor{blue}{//Redo Phase-I,II when reciprocity parameters change. }

\end{algorithmic}
\end{algorithm}

%% file: sec5_reci.tex
\section{Performance of the proposed method}
\label{sec:perf}

\subsection{ Success Probability of Eve} 

The detection accuracy of any hypothesis test (including (\ref{eq:H0H1_1_2})) is fully characterized by two kinds of detection errors: i) probability of false alarm $P_{fa}$ (declaring $H_1=1$, while in reality $H_0=1$), ii) probability of missed detection $P_{md}$ (declaring $H_0=1$, while in reality $H_1=1$). Since this work follows Neyman-Pearson procedure to compute $\delta^R$ in (\ref{eq:delta}), the detection accuracy of the test then solely depends on the success probability (probability of missed detection) of Eve: 
\begin{equation}
	\label{eq:pmd_1}
	P_{md}^R=Pr(|q^R|<\delta^R|H_1)
\end{equation}

Let $v_T=|\mu_E-\mu_A|$, $\sigma_T=\sqrt{{\Sigma_{B|E}^R}/{2}}$. Then, test statistic $T^R|H_1 \sim Rice(v_T,\sigma_T)$ and: 
\begin{equation}
	\label{eq:pmd_1}
	P_{md}^R = 1 - Q_1 ( {v_T}/{\sigma_T}, {\delta^R}/{\sigma_T} )
\end{equation}
Let $a=v_T/{\sigma_T}$, $b={\delta^R}/{\sigma_T}$. Then $Q_1(a,b)$ is the first-order Marcum Q-function:
\begin{equation}
	\label{eq:Q1}
	Q_1(a,b) = \int_b^{\infty} x \exp{\bigg( -\frac{x^2+a^2}{2} \bigg)} I_0(ax) dx
\end{equation}
where $I_0$ is the modified Bessel function of the first kind of zero order. A closed-form solution of (\ref{eq:Q1}) cannot be obtained; however, its analytical approximations do exist (see, e.g., \cite{Kam:TWC:2008}).

{\bf Remark 2.} (\ref{eq:pmd_1}) is usually solved offline, i.e., before the authentication phase commences. But solving (\ref{eq:pmd_1}) requires knowledge about Eve's fingerprint $\mu_E$ as well as $\Sigma_{B|E}^R$ (or, equivalently, knowledge of $a$ and $b$) which may not be available prior to test phase. One way to address this problem is to assume that: A1) the unknown fingerprint $\mu_E \sim \mathcal{CN}(1,1)$, A2) the distribution of $\Sigma_{B|E}^R$ (i.e., average SNR $\bar{\gamma}_{EB}$ of the link between Eve and Bob) is known to Bob\footnote{In other words, $\bar{\gamma}_{EB}$ is varied over a range (say, $0-30$ dB), and accordingly, a set of receiver operating characteristic (ROC) plots is obtained.}. With this, the approach taken is to substitute $a$ by $\mathbb{E}(a)\doteq \hat{a}$ and $b$ by $\mathbb{E}(b)\doteq \hat{b}$ in (\ref{eq:Q1}) whenever realization(s) $a$ and/or $b$ are not available:
\begin{equation}
	\label{eq:Q1_df}
	Q_1(a,b) \approx Q_1(\hat{a},\hat{b}) = \int_{\hat{b}}^{\infty} x \exp{\bigg( -\frac{x^2+\hat{a}^2}{2} \bigg)} I_0(\hat{a}x) dx
\end{equation} 

Below, two corollaries describe solution of (\ref{eq:pmd_1}) for the two cases of DF relaying and AF relaying by the sender nodes.

{\bf Corollary 2.} When sender nodes do DF relaying, due to A1, $a=v_T/{\sigma_T}\sim Rice(v_a,\sigma_a)$ where $v_a^2=(1-\frac{\mu_{A,x}}{\sqrt{\Sigma_{B|E}^{DF}/2}})^2+\frac{\mu_{A,y}^2}{\Sigma_{B|E}^{DF}/2}$, $\sigma_a^2=\frac{1}{\Sigma_{B|E}^{DF}}$ and $\mu_A=\mu_{A,x}+i\mu_{A,y}$, while $b$ is deterministic and known to Bob. Then, Bob could substitute $a$ by $\mathbb{E}(a)\doteq \hat{a}$ to compute (\ref{eq:pmd_1}) where: 
\begin{equation}
	\label{eq:mean_a}
	\mathbb{E}(a) \doteq \hat{a} = \sigma_a\sqrt{{\pi}/{2}}.L_{1/2}(-{v_a^2}/{2\sigma_a^2})
\end{equation}
where $L_{1/2}(x)=e^{x/2}\{ (1-x).I_0(-x/2)-x.I_1(-x/2) \}$ is the Laguerre polynomial, $I_0$ ($I_1$) is modified Bessel function of the first kind and zero order (first order).

{\bf Corollary 3.} When sender nodes do AF relaying, $a=v_T/{\sigma_T}$, $b={\delta^{AF}}/{\sigma_T}$ are both random variables whose distributions are difficult to find. However, note that $v_T\sim Rice(v_v,\sigma_v)$ where $v_v^2=(1-\mu_{A,x})^2+\mu_{A,y}^2$, $\sigma_v^2=1$. Also, $\sigma_T$ has the probability density function: $f_{\sigma_T}(\sigma_T,\eta,d)=2\eta. \sigma_T \exp{ \{-\eta(\sigma_T^2-d)\} }$; $\sigma_T\ge \sqrt{d}$ where $\eta=\frac{2K.P_B}{\sigma_E^2.\bar{\gamma}_{EB}}$ where $d=\frac{\sigma_B^2}{2K.P_B.\beta_{EB}^2}$. Finally, the distribution of $\delta_{AF}$ is given in Corollary 1. Then, one can make use of the following approximation (thanks to first-order Taylor expansion): $\mathbb{E}(R/S)\approx \mathbb{E}(R)/\mathbb{E}(S)$, for random variables $R$, $S$. Then:
\begin{equation}
	\label{eq:a_b}
\hat{a}=\mathbb{E}(a)\approx \mathbb{E}(v_T)/\mathbb{E}(\sigma_T) ; \hat{b}=\mathbb{E}(b)\approx \mathbb{E}(\delta^{AF})/\mathbb{E}(\sigma_T)
\end{equation}
where $\mathbb{E}(v_T)=\sigma_v\sqrt{{\pi}/{2}}.L_{1/2}(-{v_v^2}/{2\sigma_v^2})$, $\mathbb{E}(\sigma_T)=\frac{ \frac{1}{\bar{\gamma}_{EB}}. (\frac{\sigma_B^2}{\sigma_E^2.\beta_{EB}^2})+1 }{ \frac{1}{\bar{\gamma}_{EB}}. (\frac{2K.P_B}{\sigma_E^2}) } $ and $\mathbb{E}(\delta^{AF})$ is given in (\ref{eq:delta_approx}).
Then, Bob could substitute $a,b$ by $\hat{a},\hat{b}$ from (\ref{eq:a_b}) to compute (\ref{eq:pmd_1}).
\subsection{ Performance Comparison with Other Schemes} 

Table \ref{table1} provides a qualitative comparison of proposed scheme with previous schemes \cite{Xiao:TWC:2008}-\cite{Trappe:TPDS:2013}. 

\begin{table*}[b]
\centering
\caption{Qualitative comparison of Different PHY Layer Authentication Schemes }
\label{table1}
\scalebox{0.85}{
\begin{tabular}{c | c | c | c | c | c }
\hline
 {\bf Device fingerprint}  & {\bf Training needs} & {\bf Strength against forgery} & {\bf Scenario} & {\bf Mobility support} & {\bf System needs}  \\
\hline
 reciprocity (this work)  & every several secs \cite{Guillaud:SPA:2005} & high & LoS/NLoS & low-to-medium & TDD system    \\
\hline
 CFR \cite{Xiao:TWC:2008}, CIR \cite{Jitendra:COMSNETS:2010},\cite{Wang:ICC:2013}  & every channel coherence interval & high (with iid channels) & NLoS & low-to-medium & wideband system     \\ 
\hline
 RSS \cite{Trappe:TPDS:2013} & every channel coherence interval & low (Tx power attack) & NLoS & low-to-medium & none   \\
\hline   
IQ imbalance \cite{Wang:ICC:2014} & every several seconds & high & LoS/NLoS & any & none   \\
\hline  
AoA \cite{Xiong:2010:SIGCOMM} & when nodes move & high & LoS & any & multiple antennas at Rx   \\
\hline  
CFO \cite{Mahboob:ICUWB:2015},\cite{Mahboob:Globecom:2014} & one time \cite{Mahboob:Globecom:2014}, every several seconds \cite{Mahboob:ICUWB:2015} & low (freq. translate attack) & LoS/NLoS & any & none   \\
\hline  
\end{tabular}
}
\end{table*}

%% file: sec6_reci_sim.tex
\section{Numerical Results}

\label{sec:results}

\begin{figure}[ht]
\begin{center}
	\includegraphics[width=\linewidth]{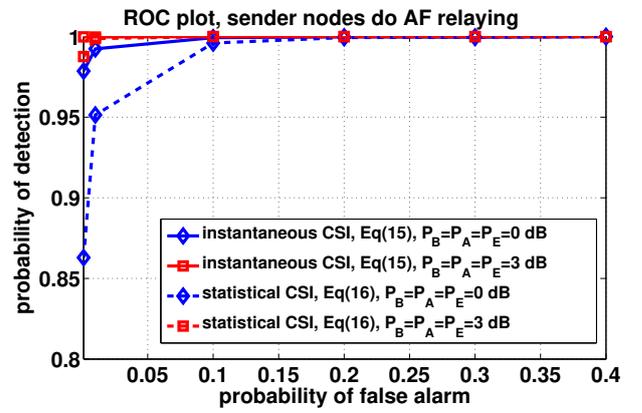}
\caption{Impact of Alice's CSI on $P_d$}
\label{f1}
\end{center}
\end{figure}

Fig. \ref{f1} plots the ROC curves for the proposed scheme when sender nodes do AF relaying and when Bob knows: i) actual realization of $\Sigma_{B|A}^{AF}$, ii) only distribution of $\Sigma_{B|A}^{AF}$. Fig.\ref{f1} suggests that the average performance loss, when Algorithm 1 operates only on statistical CSI of Alice, diminishes with increase in set-point $P_{fa}$. Fig. \ref{f1} also highlights that the detection performance of proposed scheme increases with an increase in operational SNRs $P_B,P_A,P_E$ of the system. 

\begin{figure}[ht]
\begin{center}
	\includegraphics[width=\linewidth]{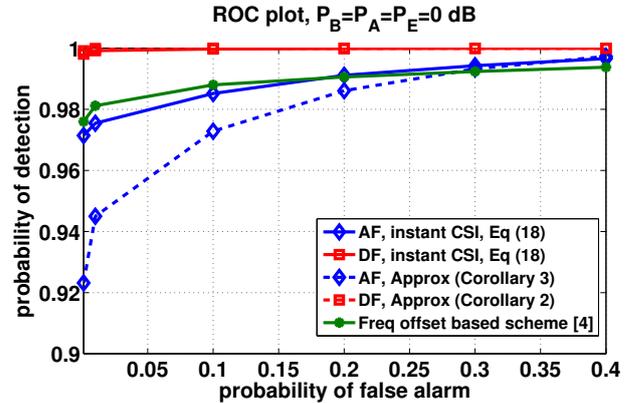}
\caption{Impact of Eve's CSI on $P_d$}
\label{f2}
\end{center}
\end{figure}

Fig. \ref{f2} plots again the ROC curves and conveys the following information: i) DF relaying by the sender nodes outperforms the AF relaying policy, ii) detection performance of the proposed scheme is comparable to that of CFO based scheme \citep{Mahboob:ICUWB:2015}, iii) the approximation in Corollary 3 is pessimistic (however, the gap between the approximation in Corollary 3 and (18) diminishes as the set-point $P_{fa}$ increases).

%% file: conclusion_reci.tex
\section{Conclusion}
\label{sec:conclusion}

We proposed a reciprocity based, sender-node authentication scheme whose detection performance is comparable to previous schemes, and is light-weight in terms of training needs. Moreover, when implementing the proposed scheme at the sender nodes, DF relaying policy is to be preferred over AF relaying policy.